\documentclass[10pt,twocolumn]{article}
\usepackage{graphicx,url,array}
\usepackage[left=1.4cm,right=1.4cm,bottom=2.4cm,top=2.1cm]{geometry}
\usepackage{flushend,caption}
\usepackage{cite}
\usepackage{amsmath,amssymb,amsfonts,bm}
\usepackage{algorithmic}
\usepackage{graphicx}
\usepackage{textcomp}
\usepackage{xcolor}
\usepackage{url}
\usepackage{multirow}
\usepackage{subfig}
\let\OLDthebibliography\thebibliography
\renewcommand\thebibliography[1]{
  \OLDthebibliography{#1}
  \setlength{\parskip}{0pt}
  \setlength{\itemsep}{6pt plus 0.3ex}
}
\def\BibTeX{{\rm B\kern-.05em{\sc i\kern-.025em b}\kern-.08em
    T\kern-.1667em\lower.7ex\hbox{E}\kern-.125emX}}
\title{WaveTransformer: A Novel Architecture for Audio Captioning Based on Learning Temporal and Time-Frequency Information}
\author{An Tran, Konstantinos Drossos, and Tuomas Virtanen\vspace{9pt}\\
Audio Research Group, Tampere University, Tampere, Finland\vspace{3pt}\\
email: \{firstname.lastname\}@tuni.fi}
\begin{document}
\date{}
\twocolumn[
\maketitle
  \begin{@twocolumnfalse}
    \maketitle
        \begin{abstract}
        \normalsize
        \noindent
        Automated audio captioning (AAC) is a novel task, where a method takes as an input an audio sample and outputs a textual description (i.e. a caption) of its contents. Most AAC methods are adapted from from image captioning of machine translation fields. In this work we present a novel AAC novel method, explicitly focused on the exploitation of the temporal and time-frequency patterns in audio. We employ three learnable processes for audio encoding, two for extracting the local and temporal information, and one to merge the output of the previous two processes. To generate the caption, we employ the widely used Transformer decoder. We assess our method utilizing the freely available splits of Clotho dataset. Our results increase previously reported highest SPIDEr to 17.3, from 16.2.\vspace{6pt}
        \end{abstract}
        \vspace{22pt}
  \end{@twocolumnfalse}
]
%
% ----------------------------------------------------------------------
% ----------------------------------------------------------------------
%
\section{Introduction}\label{sec:intro}
Automated audio captioning (AAC) is an intermodal translation task, where the system receives as an input an audio signal and outputs a textual description of the contents of the audio signal (i.e. outputs a caption)~\cite{drossos2017automated}. AAC is not speech-to-text, as the caption does not transcribe speech. In a nutshell, an AAC method learns to identify the high-level, humanly recognized information in the input audio, and expresses this information with text. Such information can include complex spatiotemporal relationships of sources and entities, textures and sizes, and abstract and high-level concepts (e.g. ``several barnyard animals mooing in a barn while it rains outside'').

There are different published approaches for AAC. Regarding input audio encoding, some approaches use recurrent neural networks (RNNs)~\cite{Wu_2019,nguyen:2020:arxiv-a,drossos2019clotho}, others 2D convolutional neural networks (CNNs)~\cite{naranjo-alcazar:2020:dcase:tech-report, koizumi2020transformerbased, wang:2020:dcase:tech-report}, and some others the Transformer~\cite{vaswani2017attention} model~\cite{shi:2020:dcase:tech-report}. Though, RNNs are known that have difficulties on learning temporal information~\cite{serdyuk:twinnet}, 2D CNNs model time-frequency but not temporal patterns~\cite{drossos2020sound}, and the Transformer was not originally designed for sequences of thousands time-steps~\cite{vaswani2017attention}. For generating the captions, the Transformer decoder~\cite{koizumi2020transformerbased,Takeuchi2020EffectsOW, shi:2020:dcase:tech-report} or RNNs~\cite{drossos2017automated,nguyen:2020:arxiv-a, naranjo-alcazar:2020:dcase:tech-report} are mostly employed, and the alignment of input audio and output captions is typically implemented with an attention mechanism~\cite{wang:2020:dcase:tech-report, Takeuchi2020EffectsOW}. Also, some approaches adopt a multi-task approach, where the AAC method is regularized by the prediction of keywords, based on the input audio~\cite{koizumi2020transformerbased,Takeuchi2020EffectsOW,cakir:2020:arxiv-a}. 

In this paper we present a novel AAC approach, based on a learnable representation of audio that is focused on encoding the information need for AAC. We adopt existing machine listening approaches where sound sources and actions are well captured by time-frequency information~\cite{drossos2020sound,cakir2017taslp}, and additionally exploit temporal information in audio using 1D dilated convolutions that operate on the time dimension~\cite{oord2016wavenet,lim2017dcase-techreport}, for learning of high-level information (e.g. background vs foreground, spatiotemporal relationships). Additionally, we claim that these two type of information can be combined, providing a well-performing learned audio representation for AAC. To this end, we present an approach which is explicitly focusing on the above aspects. 
We employ three different encoding processes for the input audio, one regarding temporal information, a second that considers the time-frequency information, and a third that merges the previous two and its output is given as an input to a decoder which generates the output caption. 

The contribution of our work is: i) we present \emph{the first method that explicitly focuses on exploiting temporal and local time-frequency information for AAC}, ii) \emph{we provide state-of-the-art (SOTA) results} using only the \emph{freely available splits of Clotho} dataset and \emph{without any data augmentation and/or multi-task learning}, and iii) we show the impact on the performance of the different components of our method, i.e. the temporal and local time-frequency information, merging the previous two, or all of them. The rest of the paper is as follows. In Section~\ref{sec:method} we present our method. Section~\ref{sec:evaluation} presents the evaluation process of our method, and the obtained results are in Section~\ref{sec:results}. Section~\ref{sec:conclusion} concludes the paper and proposes future research directions. 
%
% ----------------------------------------------------------------------
% ----------------------------------------------------------------------
%
\section{Proposed method}\label{sec:method}
\begin{figure}[t]
    \centering
    \includegraphics[trim={1.6cm 0.2cm 0.6cm 1.0cm},clip,width=\columnwidth]{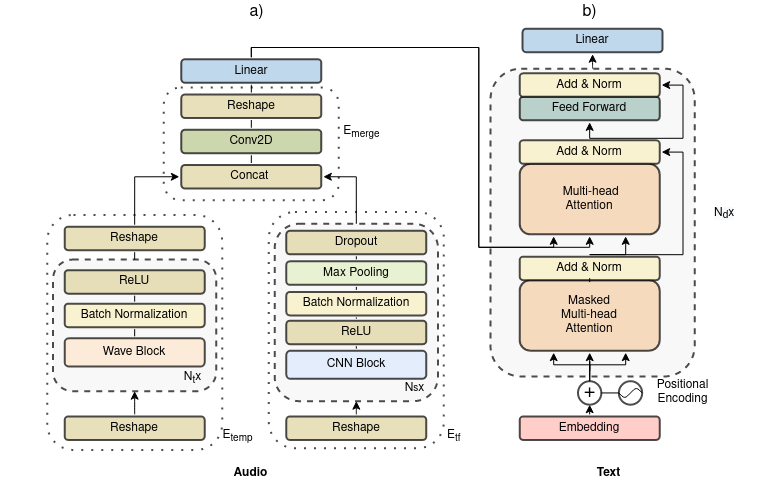}
    \caption{The WaveTransformer, with the encoder on the left-hand side and the decoder on the right-hand side}
    \label{fig:full_model}
\end{figure}

Our method takes as an input a sequence of $T_{\text{a}}$ vectors with $F$ audio features, $\mathbf{X}\in\mathbb{R}^{T_{\text{a}}\times F}$, and outputs a sequence of $T_{\text{w}}$ vectors having $W$ one-hot encoded words, $\mathbf{Y}$. To do so, our method utilizes an encoder-decoder scheme, where the encoder is based on CNNs and the decoder is based on feed-forward neural networks (FFNs) and multi-head attention. Our encoder takes $\mathbf{X}$ as an input, exploits temporal and time-frequency structures in $\mathbf{X}$, and outputs the learned audio representation $\mathbf{Z}\in\mathbb{R}^{T_{\text{a}}\times F'}$, which is a sequence of $T_{\text{a}}$ vectors of $F'$ learned audio features. The decoder takes as an input $\mathbf{Z}$ and outputs $\mathbf{Y}$. Figure~\ref{fig:full_model} illustrates our proposed method. 
%
% ----------------------------------------------------------------------
% ----------------------------------------------------------------------
%
\subsection{Encoder}\label{ssec:encoder}
Our encoder, $E(\cdot)$, consists of three learnable processes, $E_{\text{temp}}(\cdot)$, $E_{\text{tf}}(\cdot)$, and $E_{\text{merge}}(\cdot)$. $E_{\text{temp}}$ learns temporal context and frame-level information in $\mathbf{X}$~\cite{lim2017dcase-techreport}, and is inspired by WaveNet~\cite{oord2016wavenet} but with non-causal convolutions, since in AAC there is no restriction for causality in the encoding of input audio. $E_{\text{tf}}$ learns time-frequency patterns in $\mathbf{X}$, and is inspired by SOTA methods for sound event detection~\cite{drossos2020sound,cakir2017taslp}, and $E_{\text{merge}}$ merges the information extracted by $E_{\text{temp}}$ and $E_{\text{tf}}$. 

$N_{\text{t}}$ blocks of CNNs (called wave-blocks henceforth) in $E_{\text{temp}}$, sequentially process $\mathbf{X}$. Each wave-block consists of seven 1D CNNs, $\text{CNN}^{n_{\text{t}}}_{\text{t}_{1}}$ to $\text{CNN}^{n_{\text{t}}}_{\text{t}_{7}}$, with $n_{\text{t}}$ to be the index of the wave-block. For example, $\text{CNN}^{2}_{\text{t}_{3}}$ is the third CNN of the second wave-block. The kernel size, stride, and dilation of $\text{CNN}^{n_{\text{t}}}_{\{\text{t}_{1}, \text{t}_{4}, \text{t}_{7}\}}$ are one and its padding zero. The kernel size of $\text{CNN}^{n_{\text{t}}}_{\{\text{t}_{2}, \text{t}_{3}\}}$ is three and its padding, dilation, and stride is one. The kernel size of $\text{CNN}^{n_{\text{t}}}_{\{\text{t}_{5}, \text{t}_{6}\}}$ is three, its padding and dilation are two, and stride is one. $\text{CNN}^{n_{\text{t}}}_{\text{t}_{1}}$ has $C^{n_{\text{t}}}_{\text{in}}$ and $C^{n_{\text{t}}}_{\text{out}}$ input and output channels, respectively, and the rest have $C^{n_{\text{t}}}_{\text{out}}$ input and output channels. 

The above hyper-parameters are based on the WaveNet architecture~\cite{oord2016wavenet}. The output of the $n_{\text{t}}$-th wave-block, $\mathbf{H}^{n_{\text{t}}}_{\text{t}}$, is obtained by
\begin{align}
    \mathbf{H}''^{n_{\text{t}}}_{\text{t}_{1}} =& \text{CNN}^{n_{\text{t}}}_{\text{t}_{1}}(\mathbf{H}^{n_{\text{t}}-1}_{\text{t}})\text{,}\\
    \mathbf{S}''^{n_{\text{t}}}_{\text{t}} =& \tanh(\text{CNN}^{n_{\text{t}}}_{\text{t}_{2}}(\mathbf{H}''^{n_{\text{t}}}_{\text{t}_{1}}))\odot\sigma(\text{CNN}^{n_{\text{t}}}_{\text{t}_{3}}(\mathbf{H}''^{n_{\text{t}}}_{\text{t}_{1}})){,}\\
    \mathbf{H}'^{n_{\text{t}}}_{\text{t}} =& \text{CNN}^{n_{\text{t}}}_{\text{t}_{4}}(\mathbf{S}''^{n_{\text{t}}}_{\text{t}}) + \mathbf{H}''^{n_{\text{t}}}_{\text{t}_{1}}{,}\\
    \mathbf{S}'^{n_{\text{t}}}_{\text{t}} =& \tanh(\text{CNN}^{n_{\text{t}}}_{\text{t}_{5}}(\mathbf{H}'^{n_{\text{t}}}_{\text{t}}))\odot\sigma(\text{CNN}^{n_{\text{t}}}_{\text{t}_{6}}(\mathbf{H}'^{n_{\text{t}}}_{\text{t}}))\text{, and}\\
    \mathbf{H}^{n_{\text{t}}}_{\text{t}} =& \text{ReLU}(\text{BN}^{n_{\text{t}}}_{\text{t}}(\text{CNN}^{n_{\text{t}}}_{\text{t}_{7}}(\mathbf{S}'^{n_{\text{t}}}_{\text{t}}) + \mathbf{H}'^{n_{\text{t}}}_{\text{t}_{1}})){,}
\end{align}
\noindent
where $\text{BN}^{n_{\text{t}}}_{\text{t}}$ is the batch normalization process at the $n_{\text{t}}$-th wave-block, ReLU is the rectified linear unit, $\sigma(\cdot)$ is the sigmoid non-linearity, $\odot$ is the Hadamard product, $\mathbf{H}^{0}_{\text{t}}=\mathbf{X}_{\text{t}}$, and $\mathbf{H}^{N_{\text{t}}}_{\text{t}}\in\mathbb{R}^{C^{n_{\text{t}}}_{\text{out}}\times T_{\text{a}}}_{\geq0}$. The output of $E_{\text{temp}}$, $\mathbf{Z}_{\text{t}}=E_{\text{temp}}(\mathbf{X}_{\text{t}})$, is obtained by reshaping $\mathbf{H}^{N_{\text{t}}}_{\text{t}}$ to $\{1\times T_{\text{a}}\times C^{n_{\text{t}}}\}$. All $\text{CNN}^{n_{\text{t}}}$ operate along the time dimension of $\mathbf{X}_{\text{t}}$, allowing $\mathbf{H}^{N_{\text{t}}}_{\text{t}}$ to learn temporal information from $\mathbf{X}_{\text{t}}$~\cite{oord2016wavenet} and be used effectively in WaveTransformer for learning information that requires temporal context, e.g. spectro-temporal relationships. The time receptive field of each wave-block spans seven time-steps of its corresponding input, leading to a receptive field of $7N_{t}-1$ time-steps of $\mathbf{X}$, for the output of the $N_{t}$-th wave-block.

$E_{\text{tf}}$ employs $N_{tf}$ blocks of 2D CNNs, called 2DCNN-blocks henceforth. Each 2DCNN-block consists of a 2D CNN ($\text{S-CNN}^{n_{\text{tf}}}$), a leaky ReLU (LU), and a 2D CNN ($\text{P-CNN}^{n_{\text{tf}}}_{\text{tf}}$). Each 2DCNN-block is followed by a ReLU, a BN ($\text{BN}^{n_{\text{tf}}}$) process, a max-pooling ($\text{MP}^{n_{\text{tf}}}$) process that operates only on the feature dimension (hyper-parameters according to~\cite{drossos2020sound}), and a dropout (DR) with probability of $p_{n_{\text{tf}}}$. The 2DCNN-blocks are inspired by AAC and sound event detection and classification methods, and the recent, successful adoption of depth-wise separable convolutions~\cite{cakir:2020:arxiv-a,drossos2020sound,fonseca2019dcase}. The 2DCNN-blocks learn spatial time-frequency information from their input~\cite{drossos2020sound}, allowing $\mathbf{H}^{N_{\text{d}}}_{\text{d}}$ to be used effectively for the identification of sources and actions~\cite{drossos2020sound,fonseca2019dcase}.

$\text{S-CNN}^{n_{\text{tf}}}$ consists of $C^{n_{\text{tf}}}_{\text{in}}$ different (5, 5) kernels with unit stride, and padding of 2, focusing on learning time-frequency patterns from each channel of its input. Each kernel of $\text{S-CNN}^{n_{\text{tf}}}$ is applied to only one channel of the input to $\text{S-CNN}^{n_{\text{tf}}}$, according to the depthwise separable convolution model and to enforce the learning of spatial time-frequency patters~\cite{drossos2020sound}. $\text{P-CNN}^{n_{\text{tf}}}_{\text{tf}}$ consists of a square kernel of size $K_{\text{P-CNN}}>1$, with unit stride, and padding of 2, focusing on learning cross-channel information from the output of $\text{S-CNN}^{n_{\text{tf}}}$, since the kernels of $\text{P-CNN}^{n_{\text{tf}}}_{\text{tf}}$ operate on all channels of the input to $\text{P-CNN}^{n_{\text{tf}}}_{\text{tf}}$. 

While hyper-parameters of $\text{S-CNN}^{n_{\text{tf}}}$ and $\text{S-CNN}^{n_{\text{tf}}}$ are based on~\cite{drossos2020sound}, the usage of $K_{\text{P-CNN}}>1$ is not according to a typical point-wise convolution (i.e. with a (1, 1) kernel, unit stride, and zero padding), as it was experimentally found that it performs better, using the training and validation data, and the protocol described in Section 3. $\text{S-CNN}^{1}$ has $C^{n_{\text{tf}}}_{\text{in}}=1$ and $C^{n_{\text{tf}}}_{\text{out}}=C^{n_{\text{t}}}_{\text{out}}$ input and output channels, respectively. $\text{S-CNN}^{n_{\text{tf}}>1}$ and $\text{P-CNN}^{n_{\text{tf}}}$ have input and output channels equal to $C^{n_{\text{t}}}_{\text{out}}$. The output of the $n_{\text{tf}}$-th 2DCNN-block, $\mathbf{H}^{n_{\text{tf}}}_{\text{tf}}\in\mathbb{R}^{C^{n_{\text{tf}}}_{\text{out}}\times T_{\text{a}}\times F'_{\text{tf}}}_{\geq0}$, is obtained by
\begin{align}
    \mathbf{S}'^{n_{\text{tf}}}_{\text{tf}} =& \text{P-CNN}^{n_{\text{tf}}}(\text{BN}^{n_{\text{tf}}}(\text{LU}(\text{S-CNN}^{n_{\text{tf}}}(\mathbf{H}^{n_{\text{tf}}-1}_{\text{tf}}))))\text{ and}\\
    \mathbf{H}^{n_{\text{tf}}}_{\text{tf}} =& \text{DR}(\text{MP}^{n_{\text{tf}}}(\text{BN}^{n_{\text{tf}}}(\mathbf{S}'^{n_{\text{tf}}}_{\text{tf}})))\text{, }
\end{align}
\noindent
where $\mathbf{H}^{0}_{\text{tf}}=\mathbf{X}_{\text{tf}}$ and $\mathbf{H}^{N_{\text{tf}}}_{\text{tf}}\in\mathbb{R}^{C^{N_{\text{tf}}}_{\text{out}}\times T_{\text{a}}\times 1}_{\geq0}$. Then, $\mathbf{Z}_{\text{tf}}=E_{\text{tf}}(\mathbf{X}_{\text{tf}})$ is obtained by reshaping $\mathbf{H}^{N_{\text{tf}}}_{\text{tf}}$ to $\{1\times T_{\text{a}}\times C^{N_{\text{tf}}}_{\text{out}}\}$. 

$E_{\text{merge}}$ consists of a 2D CNN, $\text{CNN}_{\text{m}}$ and a feed-forward neural network (FNN), $\text{FNN}_{\text{m}}$, with shared weights through time. Specifically, $\text{CNN}_{\text{m}}$ has a (5, 5) kernel with unit stride and dilation, padding of 2, and two input and one output channels. Both $\mathbf{Z}_{\text{t}}$ and $\mathbf{Z}_{\text{tf}}$ have the same dimensionality, are concatenated in their channel dimension, and given as an input to $\text{CNN}_{\text{m}}$, as 
$\mathbf{Z}'' = [\mathbf{Z}_{\text{t}};\mathbf{Z}_{\text{tf}}]$ and $\mathbf{Z}' = \text{CNN}_{\text{m}}(\mathbf{Z}'')$, where $\mathbf{Z}''\in\mathbb{R}^{2\times T_{\text{a}}\times C^{N_{\text{tf}}}_{\text{out}}}_{\geq0}$, and $\mathbf{Z}'\in\mathbb{R}^{1\times T_{\text{a}}\times C^{N_{\text{tf}}}_{\text{out}}}$ is the output of $\text{CNN}_{\text{m}}$. $\mathbf{Z}'$ is then reshaped to $\{T_{\text{a}}\times C^{N_{\text{tf}}}_{\text{out}}\}$ and given as an input to $\text{FNN}_{\text{m}}$, as
$\mathbf{Z} = \text{FNN}_{\text{m}}(\mathbf{Z}')\text{,}$ where $\mathbf{Z}\in\mathbb{R}^{T_{\text{a}}\times F'}$, with $F'=C^{N_{\text{tf}}}_{\text{out}}$.
%
% ----------------------------------------------------------------------
% ----------------------------------------------------------------------
%
\subsection{Decoder}\label{ssec:decoder}
We employ the decoder of the Transformer model~\cite{vaswani2017attention} as our encoder, $D(\cdot)$. During training $D$ takes as an input $\mathbf{Y}$ and $\mathbf{Z}$, and outputs a sequence of $T_{\text{w}}$ vectors having a probability distribution over $W$ words, $\hat{\mathbf{Y}}\in[0,1]^{T_{\text{w}}\times W}$. We follow the implementation in~\cite{vaswani2017attention}, employing an FFN as embedding extractor for one-hot encoded words, $\text{FNN}_{\text{emb}}(\cdot)$, a positional encoding process, $P_{\text{enc}}(\cdot)$, $N_{\text{dec}}$ decoder blocks, $D^{n_{\text{dec}}}(\cdot)$, and an FFN at the end which acts as a classifier, $\text{FNN}_{\text{cls}}(\cdot)$. $\text{FNN}_{\text{emb}}$ and $\text{FNN}_{\text{cls}}$ have their weights shared across the words of a caption. Each $D^{n_{\text{dec}}}$ consists of a masked multi-head self-attention, a layer-normalization (LN) process, another multi-head attention that attends at $\mathbf{Z}$, followed by another LN, an FNN, and another LN.

We model each $D^{n_{\text{dec}}}$ as a function taking two inputs, $\mathbf{U}^{n_{\text{dec}}}\in\mathbb{R}^{T_{\text{w}}\times V^{n_{\text{dec}}}_{\text{e}}}$ and $\mathbf{Z}$, and having as output $\mathbf{H}^{n_{\text{dec}}}_{\text{dec}}\in\mathbb{R}^{T_{\text{w}}\times V^{n_{\text{dec}}}_{\text{e}}}$, with $\mathbf{H}^{0}_{\text{dec}}=\mathbf{H}'_{\text{dec}}$, $\mathbf{U}^{0}=\mathbf{Y}$, and $V^{0}_{\text{e}}=W$. All FNNs of each $D^{n_{\text{dec}}}$ have input-output dimensionality of $V^{n_{\text{dec}}}_{\text{e}}$. We use $N_{\text{att}}$ attention heads and for the multi-head attention layers and $p_{\text{d}}$ dropout probability. For the implementation details, we refer the reader to the paper of Transformer model~\cite{vaswani2017attention}. $\text{FNN}_{\text{emb}}$ takes as an input $\mathbf{Y}$ and its output is processed by the positional encoding process, as
\begin{equation}
    \mathbf{H}'_{\text{dec}} = P_{\text{enc}}(\text{FNN}_{\text{emb}}(\mathbf{Y})))\text{,}
\end{equation}
\noindent
where $P_{\text{enc}}$ is according to the original paper~\cite{vaswani2017attention}. $\mathbf{H}'_{\text{dec}}$ is processed serially by the $N_{\text{dec}}$ decoder blocks, as $\mathbf{H}^{n_{\text{dec}}}_{\text{dec}} = D^{n_{\text{dec}}}(\mathbf{H}^{n_{\text{dec}}-1}_{\text{dec}},\mathbf{Z})$, and then we obtain $\hat{\mathbf{Y}}$ as
\begin{equation}
    \hat{\mathbf{Y}} = \text{FNN}_{\text{cls}}(\mathbf{H}^{N_{\text{dec}}}_{\text{dec}})\text{.}
\end{equation}
\noindent
We optimize jointly the parameters of the encoder and decoder, by minimizing the cross-entropy loss between $\mathbf{Y}$ and $\hat{\mathbf{Y}}$.
%
% ----------------------------------------------------------------------
% ----------------------------------------------------------------------
%
\begin{table*}[!t]
    \centering
    \caption{Results on Clotho evaluation dataset. B$_{n}$ stands for BLEU$_{n}$. Boldface fonts indicate the best values for each metric}
    \label{tab:results}
    \resizebox{\textwidth}{!}{%
    \begin{tabular}{l*{9}{c}}
        \textbf{Model} & \textbf{B\textsubscript{1}} & \textbf{B\textsubscript{2}} & \textbf{B\textsubscript{3}} & \textbf{B\textsubscript{4}} & \textbf{METEOR} & \textbf{ROUGE\textsubscript{L}} & \textbf{CIDEr} & \textbf{SPICE} & \textbf{SPIDEr}\\
        \hline
        TRACKE (w/o MT)~\cite{koizumi2020transformerbased} & 50.2 & 29.9 & 18.3 & 10.2 & 14.1 & \textbf{33.7} & 23.3 & 09.1 & 16.2\\
        NTT (w/o MT, DA, and PP)~\cite{Takeuchi2020EffectsOW} & \textbf{52.1} & 29.4 & 17.4 & 10.3 & 13.8 & 33.5 & 23.2 & 08.5 & 15.8\\
        NTT (MT+PP, w/o DA)~\cite{Takeuchi2020EffectsOW} & 52.0 & \textbf{31.2} & \textbf{20.0} & \textbf{12.7} & 14.0 & 33.7 & 26.1 & 08.2 & 17.2\\
        WT\textsubscript{temp} & 45.8 & 25.9 & 15.4 & 08.8 & 13.9 & 32.0 & 19.8 & 08.7 & 14.2\\
        WT\textsubscript{tf} & 47.9 & 28.0 & 17.1 & 10.2 & 14.7 & 33.1 & 24.7 & 09.3 & 17.0 \\
        WT\textsubscript{avg}  & 47.9 & 28.1 & 17.1 & 10.3 & 14.8 & 33.0 & 24.7 & 09.4 & 17.0 \\ 
        WT & 48.4 & 28.2 & 17.4 & 10.2 & \textbf{14.8} & 33.2 & 24.7 & \textbf{09.9} & 17.3\\
        WT-B & 49.8 & 30.3 & 19.7 & 12.0 & 14.3 & 33.2 & \textbf{26.8} & 09.5 & \textbf{18.2}
    \end{tabular}
    }
\end{table*}
%
% ----------------------------------------------------------------------
% ----------------------------------------------------------------------
%
\section{Evaluation}\label{sec:evaluation}
To evaluate our method, we employ the dataset and protocol defined at the AAC task at the DCASE2020 challenge. The code and the pre-trained weights of our method are freely available online\footnote{\url{https://github.com/haantran96/wavetransformer}}. We also provide an online demo of our method, with 10 audio files, the corresponding predicted captions, and the corresponding ground truth captions\footnote{\url{https://haantran96.github.io/wavetransformer-web-demo/}}. 

\subsection{Dataset and pre- and post-processing}\label{ssec:data-prorpcessing}
We employ the freely available and well curated AAC dataset, Clotho, consisting
of around 5000 audio samples of CD quality, 15 to 30 seconds long, and each sample is annotated by human annotators with five captions of eight to 20 words, amounting to around 25 000 captions~\cite{drossos2019clotho,lipping2019dcase}. Clotho is divided in three splits: i) development, with 14465 captions, ii) evaluation, with 5225, and iii) testing with 5215 captions. We employ development and evaluation splits which are publicly and freely available. We extract $F=64$ log mel-band energies using Hamming window of 46ms with 50\% overlap from the audio files, resulting to $1292\leq T_{\text{a}}\leq2584$, for audio samples which length is between 15 and 30 seconds.

We process each caption and we prepend and append the \textless sos\textgreater~(start-of-sentence) and \textless eos\textgreater~(end-of-sentence) tokens, respectively. Additionally, we process the development split and we randomly select and reserve 100 audio samples and their captions in order to be used as a validation split during training. These 100 samples are selected according to the criterion that their captions do not contain a word that appears in the captions of less than 10 audio samples. We term the resulting training (i.e. development minus the 100 audio samples) and validation splits as Dev\textsubscript{tra} and Dev\textsubscript{val}, respectively. We also provide the file names from Clotho development split used in Dev\textsubscript{val}, at the online repository of WaveTransformer\textsuperscript{2}. We post-process the output of WaveTransformer during inference, employing both greedy and beam search decoding. Greedy decoding stops when \textless eos\textgreater~token or when 22 words are generated. 
%
% ----------------------------------------------------------------------
% ----------------------------------------------------------------------
%
\subsection{Hyper-parameters, training, and evaluation}
We employ the Dev\textsubscript{tra} (as training split) and Dev\textsubscript{val} (as validation split) to optimize the hyper-parameters of our method, using an early stopping policy with a patience of 10 epochs. We employ Adam optimizer~\cite{kingma2014adam}, a batch size of 12, and clipping of the 2-norm of the gradients to the value of 1. The employed hyper-parameters of our method are $N_{\text{t}}=4$, $N_{\text{tf}}=3$, $C^{n_{\text{t}}}_{out}=V_{\text{e}}=128$, $F'_{tf}=1$, $N_{\text{dec}}=3$, $N_{\text{att}}=4$, $p_{n_{\text{tf}}}=p_{\text{d}}=0.25$, and beam size of 2. This leads to the modelling of $7N_{\text{t}}-1=27$ frames, equivalent to 0.7 seconds for current $\mathbf{X}$, for $E_{\text{temp}}$. 

To assess the performance of WaveTransformer (WT) and the impact of $E_{\text{temp}}$, $E_{\text{tf}}$, $E_{\text{merge}}$, and beam search, we employ the WT, WT without $E_{\text{tf}}$ and $E_{\text{merge}}$ ($\text{WT}_{\text{temp}}$), without $E_{\text{temp}}$ and $E_{\text{merge}}$ ($\text{WT}_{\text{tf}}$), and without $E_{\text{merge}}$ ($\text{WT}_{\text{avg}}$), where we replace $E_{\text{merge}}$ with an average between $E_{\text{temp}}$ and $E_{\text{tf}}$. We evaluate the performance of WT with greedy decoding and with beam searching (indicated as WT-B) on Clotho evaluation split and using the machine translation metrics BLEU$_{1}$ to BLEU$_{4}$ scores, METEOR, and ROUGE\textsubscript{L}~\cite{papineni2002bleu,lavie2007meteor,lin2004rouge}, and the captioning metrics CIDEr, SPICE, and SPIDEr~\cite{vedantam2015cider,anderson2016spice,2017:liu:iccv}. In a nutshell, BLEU$_{n}$ measures a weighted geometric mean of modified precision of $n$-grams, METEOR measures a harmonic mean of recall and precision for segments between the two captions, and ROUGE\textsubscript{L} calculates an F-measure using the longest common sub-sequence. On the other hand, CIDEr calculates a weighted cosine similarity of $n$-grams, using term-frequency inverse-document-frequency weighting, SPICE measures how well the predicted caption recovers objects, attributes, and their relationships, and SPIDEr is the average of CIDEr and SPICE, exploiting the advantages of CIDEr and SPICE. 

Additionally, we compare our method with the two highest-performing AAC methods, NTT~\cite{Takeuchi2020EffectsOW} and TRACKE~\cite{koizumi2020transformerbased}, developed and evaluated using only Clotho development and evaluation splits. NTT uses different components, like multi-task learning (MT), data augmentation (DA), and post-processing (PP), but authors provide results without these components. TRACKE is the current SOTA, it also uses MT but the authors provide results without MT. We compare our WT against TRACKE without MT and NTT without (w/o) DA. 
%
% ----------------------------------------------------------------------
% ----------------------------------------------------------------------
%
\section{Results}
\label{sec:results}
Table~\ref{tab:results} presents the results of WT, NTT, and TRACKE. As can be seen, the learning of time-frequency information (WT\textsubscript{tf}) can lead to better results than learning temporal information (WT\textsubscript{temp}) instead. We hypothesize that this because the decoder can learn an efficient language model, filling the connecting gaps (e.g. interactions of objects) between sound events learned from $E_{\text{tf}}$. Though, from the results can be seen that employing both $E_{\text{temp}}$ and $E_{\text{tf}}$ increases more the performance of the WaveTransformer (WT). 

Comparing the different scores for the employed metrics and for the WT\textsubscript{tf} and WT cases, can be seen that the utilization of $E_{\text{temp}}$ is not contributing much in the ordering of words, as indicated by the difference of BLEU metrics between WT\textsubscript{tf} and WT. We can see that with the $E_{\text{temp}}$, our method learns better attributes of objects and their relationships, as indicated by CIDEr and SPICE scores. Thus, we argue that $E_{\text{temp}}$ contributes in learning attributes and interactions of objects, while $E_{\text{tf}}$ contributes information about objects and actions (e.g. sound events). Also, by observing the results for WT\textsubscript{avg}, we can see that a simple averaging of the learned information by $E_{\text{temp}}$ and $E_{\text{tf}}$ leads to a better description of objects, attributes, and their relationships (indicated by SPICE). Though, as can be seen by comparing WT\textsubscript{avg} and WT, the $E_{\text{merge}}$ manages to successfully merge the information by $E_{\text{temp}}$ and $E_{\text{tf}}$. The utilization of beam search (WT-B) gives a significant boost to the performance, reaching up to 18.2 SPIDEr. Compared to TRACKE and NTT methods, we can see that when excluding DA, MT, and PP, our method (WT) performs better. Additionally, WT-B performs better than NTT with MT and PP. Our post-processing consists only on using beam search, where the NTT method involves a second post-processing technique by augmenting the input data and averaging the predictions. Thus, WT surpasses NTT and TRACKE methods, setting the new SOTA of AAC. 

Finally, two, high SPIDEr-scoring, captions are for the files \emph{Flipping pages.wav}, and \emph{110422\_village\_dusk.wav} of the evaluation split of Clotho. Our predicted captions for each of these files, using WT-B, are: ``a person is flipping through the pages of a book'' and ``a dog is barking while birds are chirping in the background'', respectively, and the best matching ground truth captions are ``a person is flipping through pages in a notebook'' and ``a dog is barking in the background while some children are talking and birds are chirping'', respectively.  
%
% ----------------------------------------------------------------------
% ----------------------------------------------------------------------
%
\section{Conclusion}\label{sec:conclusion}
In this paper we presented a novel architecture for AAC, based on convolutional and feed-forward neural networks, called WaveTransformer (WT). WT focuses on learning long temporal and time-frequency information from audio, and expressing it with text using the decoder of the Transformer model. We evaluated WT using the dataset and the metrics adopted in the AAC DCASE Challenge, and we compared our method against previous SOTA methods and the DCASE AAC baseline. The obtained results show that learning time-frequency information, combined with a good language model, can lead to good AAC performance, but incorporating long temporal information can boost the obtained scores. 
%
% ----------------------------------------------------------------------
% ----------------------------------------------------------------------
%
\section*{Acknowledgement}
The authors wish to thank D. Takeuchi and Y. Koizumi for their input on previously reported results, and to acknowledge CSC-IT Center for Science, Finland, for computational resources. Part of the need computations was implemented on a GPU donated from NVIDIA to K. Drossos.
%
% ----------------------------------------------------------------------
% ----------------------------------------------------------------------
%
\bibliographystyle{IEEEbib}
\bibliography{strings,refs}
\end{document}